\documentclass[runningheads]{llncs}

\usepackage{caption}
\usepackage{subfigure}
\usepackage[T1]{fontenc}

\usepackage{cite}
\usepackage{amsmath,amssymb,amsfonts}
\usepackage{algorithmic}
\usepackage{graphicx}
\usepackage{textcomp}
\usepackage{xcolor}
\usepackage{color, colortbl}
\definecolor{Gray}{gray}{0.9}
\usepackage{amsmath}
\usepackage{pifont}

\usepackage{makecell}
\usepackage{wrapfig}
\usepackage{picinpar}

\usepackage[utf8]{inputenc} %
\usepackage[hidelinks]{hyperref}       %
\usepackage{url}            %
\usepackage{booktabs}       %
\usepackage{amsfonts}       %
\usepackage{nicefrac}       %
\usepackage{microtype}      %

\usepackage{algorithmic}
\usepackage{graphicx}
\usepackage{textcomp}
\usepackage{xcolor}
\usepackage{multirow}
\usepackage{algorithm}
\usepackage{bm}
\usepackage{subfigure}
\usepackage{caption}

\usepackage{CJKutf8}
\usepackage[T1]{fontenc}
\usepackage{graphicx}
\begin{document}
\title{CaseGNN: Graph Neural Networks for Legal Case Retrieval with Text-Attributed Graphs}
\titlerunning{CaseGNN}
\author{Yanran Tang \and Ruihong Qiu \and Yilun Liu \and Xue Li \and Zi Huang}
\authorrunning{Y. Tang et al.}
\institute{The University of Queensland\\
\email{\{yanran.tang, r.qiu, yilun.liu, helen.huang\}@uq.edu.au, xueli@eecs.uq.edu.au}}

\maketitle              %
\begin{abstract}
Legal case retrieval is an information retrieval task in the legal domain, which aims to retrieve relevant cases with a given query case. Recent research of legal case retrieval mainly relies on traditional bag-of-words models and language models. Although these methods have achieved significant improvement in retrieval accuracy, there are still two challenges: (1) \textbf{Legal structural information neglect}. Previous neural legal case retrieval models mostly encode the unstructured raw text of case into a case representation, which causes the lack of important legal structural information in a case and leads to poor case representation; (2) \textbf{Lengthy legal text limitation}. When using the powerful BERT-based models, there is a limit of input text lengths, which inevitably requires to shorten the input via truncation or division with a loss of legal context information. In this paper, a graph neural networks-based legal case retrieval model, CaseGNN, is developed to tackle these challenges. To effectively utilise the legal structural information during encoding, a case is firstly converted into a Text-Attributed Case Graph (TACG), followed by a designed Edge Graph Attention Layer and a readout function to obtain the case graph representation. The CaseGNN model is optimised with a carefully designed contrastive loss with easy and hard negative sampling. Since the text attributes in the case graph come from individual sentences, the restriction of using language models is further avoided without losing the legal context. Extensive experiments have been conducted on two benchmarks from COLIEE 2022 and COLIEE 2023, which demonstrate that CaseGNN outperforms other state-of-the-art legal case retrieval methods. The code has been released on \url{https://github.com/yanran-tang/CaseGNN}.

\end{abstract}
\section{Introduction}
Legal case retrieval (LCR) is a specialised and indispensable retrieval task that focuses on retrieving relevant cases given a query case. For legal practitioners such as judges and lawyers, using retrieval models is more efficient than manually finding relevant cases by looking into thousands of legal documents. It is said that 59\% of lawyers in the US are using web-based software to get technical services and solution suggestions\footnote{\url{https://www.clio.com/blog/lawyer-statistics/}}. LCR also greatly helps a broader community who has legal questions but does not want to spend money on expensive consultation fees.

Existing LCR models can be categorised into two streams: statistical models and language models (LM). Statistical models~\cite{BM25,TF-IDF,LMIR} focus on measuring term frequency as the case similarity while LMs~\cite{Law2Vec,Lawformer,MTFT-BERT,MVCL,BERT-PLI,LEGAL-BERT,SAILER,JOTR,DoSSIER,RPRS, NOWJ,ConversationalAgent,UA@COLIEE2022,CL4LJP} conduct nearest neighbour search with case representations from language models. Among the LMs for LCR task, there are different case text matching strategies that focus on sentence~\cite{IOT-Match}, paragraph~\cite{BERT-PLI} and whole-case~\cite{SAILER} levels.

Although the powerful LMs have achieved higher accuracy performance compared to traditional statistical models in LCR, two critical challenges still remain unsolved. (1) \textbf{Legal structural information neglect}. Under the context of legal domain, the case structural information typically refers to the relationship among different elements in a legal case, such as parties, crime activities and evidences. Recent LMs for LCR are trying to encode the unstructured raw text of a legal case into a high dimensional representation to measure the similarity with different text matching strategies~\cite{BERT-PLI,SAILER,IOT-Match}. However, only using bag-of-words statistical retrieval models or sequence LMs will restrict the interactions between different elements of cases, which will cause a significant loss of the useful structural information of legal cases. (2) \textbf{Lengthy legal text limitation}. As in the study of the LCR benchmark, COLIEE2023~\cite{COLIEE2023}, the average length of a case is 5,566 tokens~\cite{promptcase}, which exceeds most input limitations of LMs, such as 512-token limit of BERT-based models~\cite{BERT}. Therefore, most existing researches rely on truncation~\cite{SAILER} or division~\cite{BERT-PLI} to shorten the input text. These pre-processing methods for adapting the LMs' input length will lead to incomplete legal case text and finally cause the loss of legal information from a global view.  

To address the above two challenges, a novel CaseGNN framework is proposed in this paper. Firstly, for each case, the informative and useful case structural information that refers to the parties, crime activities or evidences of cases will be extracted by Named Entity Recognition and Relation Extraction tools to construct a case graph. Furthermore, to collaboratively utilise the textual and structural information in legal cases, the extracted structural information will be represented by LMs to transform the case graph into a  Text-Attributed Case Graph (TACG). Secondly, to effectively obtain a case representation from the TACG, a CaseGNN framework is proposed by utilising the Edge Graph Attention Layer (EdgeGAT) and a readout function to obtain a graph level representation for retrieval. Finally, to train CaseGNN, a contrastive loss is designed to incorporate effective easy and hard negative samples. Empirical experiments are conducted on two benchmark datasets COLIEE 2022~\cite{COLIEE2022} and COLIEE 2023~\cite{COLIEE2023}, which demonstrates that the proposed case structural information and CaseGNN can achieve state-of-the-art performance on LCR by effectively leveraging the structural information. The main contributions of this paper are summarised as follows:
\begin{itemize}
\item A CaseGNN framework is proposed for LCR to tackle the challenges on incorporating legal structural information and avoiding overlong input text.
\item A  Text-Attributed Case Graph (TACG) is developed to transform the format of unstructured case text into structural and textual format.
\item A GNN layer called edge graph attention layer (EdgeGAT) is designed to learn the representation in the TACG.
\item Extensive experiments conducted on two benchmark datasets demonstrate the state-of-the-art performance of CaseGNN.
\end{itemize}

\section{Related Work}
\subsection{Legal Case Retrieval Models}
As a specialised information retrieval (IR) task, the methods of LCR task can be categorised into two streams as IR task: statistical retrieval models~\cite{TF-IDF,BM25,LMIR} and language models~\cite{ColBERT,Doc2query,Sentence-BERT,crossencoder}. In LCR task, TF-IDF~\cite{TF-IDF}, BM25~\cite{BM25} and LMIR~\cite{LMIR} are the statistical models that also frequently used, which are all based on calculating the text matching score by utilising term frequency and inverse document frequency of words in legal case. General LMs~\cite{DeepCT,BERT,RoBERTa,monot5} are highly used for LCR task by using LMs to encode the case into representative embeddings for the powerful language understanding ability of LMs~\cite{Law2Vec,Lawformer,MTFT-BERT,MVCL,LEGAL-BERT,JOTR,DoSSIER,RPRS,NOWJ,UA@COLIEE2022,IOT-Match,Law-Match,LEDsummary,BM25injtct,LeiBi,LEVEN,JNLP@COLIEE2019,CL4LJP,QAjudge,query_conversational_agent, ConversationalAgent}. In the state-of-the-art research, to tackle the long text problem in legal domain, BERT-PLI~\cite{BERT-PLI} divides cases into paragraphs and calculates the similarity between two paragraphs while SAILER~\cite{SAILER} directly truncates the case text to cope with the input limit of LM. 

\subsection{Graph Neural Networks}
GNN models can effectively capture the structural information from graph data~\cite{GCN,GAT,GraphSAGE,fgnn,fgnnj,gag,cat}. To further utilise the edge information, SCENE~\cite{SCENE} proposed a GNN layer that can deal with the edge weights. Recently, text-attributed graph are widely used for the capacity of combining both the text understanding ability of LMs and the structural information of graphs, such as TAPE~\cite{TAPE}, G2P2~\cite{G2P2} and TAG~\cite{TAG}. 

There are two existing graph-based legal understanding methods, LegalGNN for legal case recommendation~\cite{LegalGNN} and SLR for LCR~\cite{SLR}. Both methods utilise an external legal knowledge database, such as legal concepts and charges, to construct a knowledge graph with human knowledge while encoding legal cases with general LMs. Our proposed CaseGNN is different from these two methods that there is no external knowledge and the encoding of a case actually uses the structural information from the case itself.

\section{Preliminary}
In the following, a bold lowercase letter denotes a vector, a bold uppercase letter denotes a matrix, a lowercase letter denotes a scalar or a sequence of words, and a scripted uppercase letter denotes a set.

\subsection{Task Definition}
\label{sec: task}
In legal case retrieval, given the query case $q$, and the set of $n$ candidate cases $D=\{d_1,d_2,...,d_n\}$, our task is to retrieve a set of relevant cases $D^* = \{d^*_i| d^*_i \in D \wedge relevant (d^*_i, q) \}$, where $relevant (d^*_i, q)$ denotes that $d^*_i$ is a relevant case of the query case $q$. The relevant cases are called precedents in legal domain, which refer to the historical cases that can support the judgement of the query case.

\subsection{Graph Neural Networks}
\paragraph{Graph:} A graph is denoted as $G = (V, E)$, where a node $v$ with feature $\textbf{x}_v \in \mathbb{R}^d$ for $v \in V$, and an edge with feature $\textbf{e}_{uv} \in \mathbb{R}^d$ for $e \in E$ between node $u$ and $v$.

\paragraph{Graph Neural Networks:} GNNs utilise node features, edge features, and the graph structure to learn representations for nodes, edges and the graph. Most GNNs use iterative neighbourhood aggregation to calculate the representations. After $l-1$ iterations of aggregation, the output features of a node $v$ after $l$-th layer is:
\begin{equation}
    \textbf{h}_v^{l}=\text{Map}^{l} (\textbf{h}_v^{l-1}, \text{Agg
    }({\textbf{h}_v^{l-1}, \textbf{h}_u^{l-1}, \textbf{h}_{e_{uv}}):u \in N(v)})),
\end{equation}
where $\textbf{h}_v^{l} \in \mathbb{R}^d$ is the node representation of $v$ at $l$th layer, $\textbf{h}_{e_{uv}} \in \mathbb{R}^d$ is the edge representation between node $u$ and $v$, and $N(v)$ is the neighbour node set of node $v$. Specially, the input of the first layer is initialised as $\textbf{h}_v^{0} = \textbf{x}_v$. Agg and Map are two functions that can be formed in different ways, where Agg performs aggregation to the neighbour node features and edge features while Map utilises the node self features and the neighbour features together for mapping node $v$ to a new feature vector. To generate a graph representation $\textbf{h}_G$, a Readout function is used to transform all the node features:
\begin{equation}
    \textbf{h}_G=\text{Readout}(G).
\end{equation}

\section{Method}

\begin{figure}[!t]
\centering
\includegraphics[width=\textwidth]{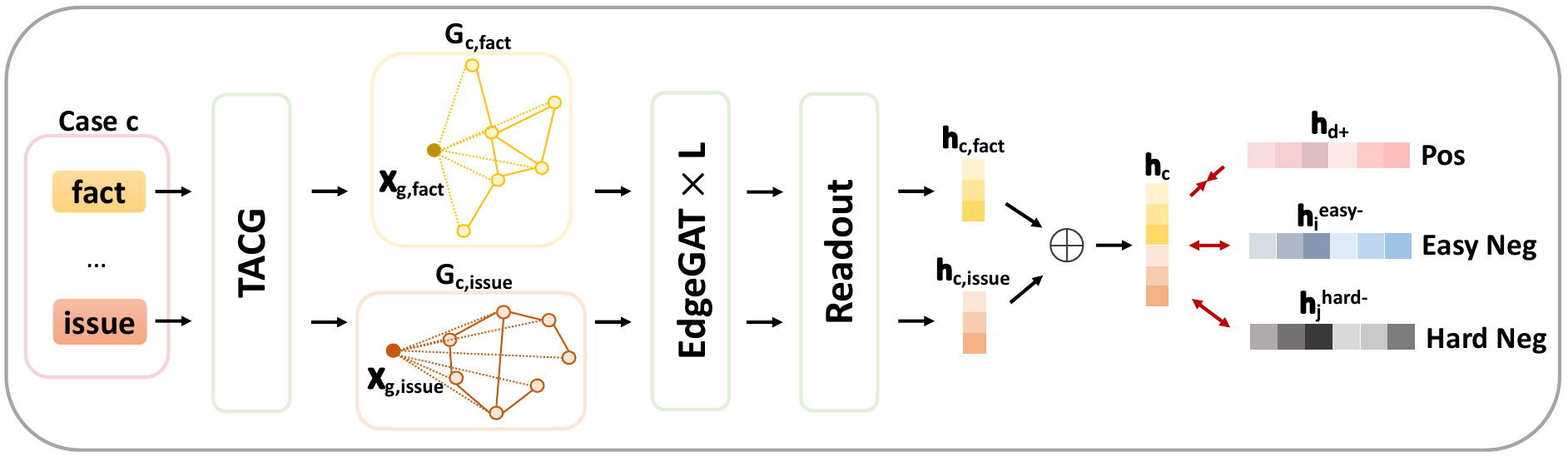}
\caption{The framework of CaseGNN. Given legal case $c$, the legal fact and the legal issue sections are converted into TACG based on information extraction and text encodings. The TACG is processed by L layers of EdgeGAT and a Readout function to obtain an overall case graph representation. The whole framework is trained with the contrastive loss with positive and negative samples.}
\label{fig:CaseGNN}
\end{figure}

\subsection{Text-Attributed Case Graph}
\label{sec:TACG}
Text-Attributed Case Graph (TACG) aims to convert the unstructured case text into a graph. To construct a TACG, the structure and the features of the graph will be obtained by using information extraction tools and language models.

\subsubsection{Information Extraction.}
\label{sec:extract}
To leverage the legal structural information for graph construction, named entity recognition tool and relation extraction tool are used for information extraction. From the legal perspective, the determining factor of relevant cases is the alignment of \textit{legal fact} and \textit{legal issue}~\cite{promptcase}. Specifically, legal fact is a basic part of a case that describe ``who, when, what, where and why'' while legal issue is the legal disputes between parties of a case and need to be settled by judges~\cite{promptcase}. The details of generating legal fact and legal issue can be found in PromptCase~\cite{promptcase}. Therefore, in this paper, the important legal structural information refers to the relation triplets that generated from legal fact and legal issue, which are extracted from a case text using the PromptCase framework~\cite{promptcase}. For example, in COLIEE datasets collected from the federal court of Canada~\cite{COLIEE2022,COLIEE2023}, a triplet example is extracted as ($applicant$, $is$, $Canadian$) from a sentence ``The applicant is a Canadian.'' in legal fact of a case, where $applicant$ denotes the ``who'' and $is, Canadian$ refers to the ``what'' in the case. After conducting information extraction, a set of triplets $R\ = \{(h,\ r,\ t)_{i=1:n}\}$ can be obtained, where $h$ is the head entity, $t$ is the tail entity, $r$ is the relation between $h$ and $t$, and $n$ is the number of triplets in a case.

\begin{wrapfigure}{r}{0.65\linewidth}
    \vspace{-0.6cm}
    \centering
    \includegraphics[width=1\linewidth]{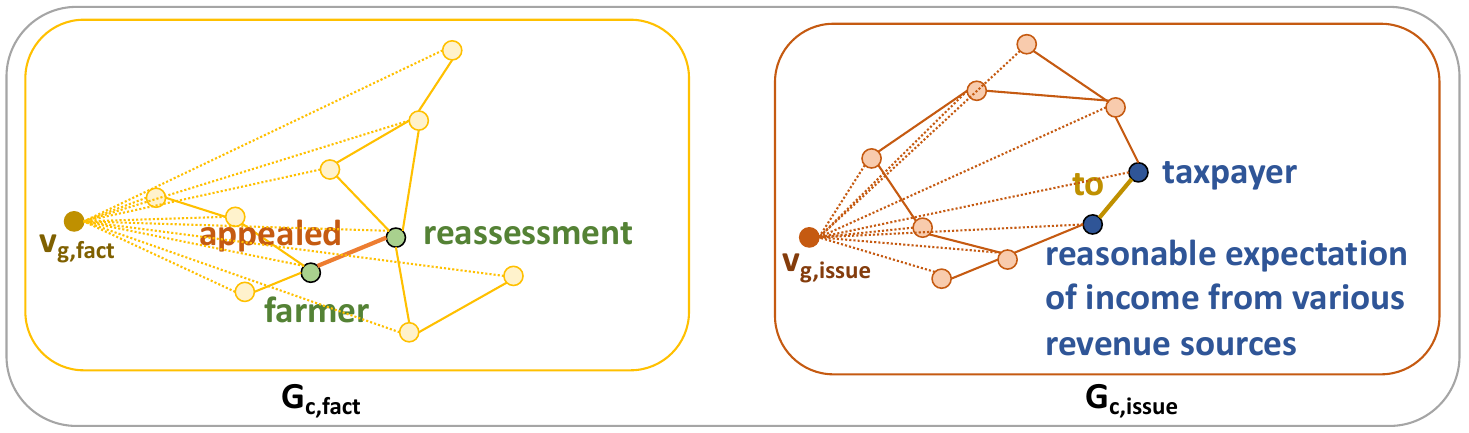}
    \caption{The constructed legal fact graph, $G_{c,\text{fact}}$ and legal issue graph, $G_{c,\text{issue}}$ of case $c$.}
    \vspace{-0.6cm}
    \label{fig:TACG}
\end{wrapfigure}

\subsubsection{Graph Construction.}
\label{sec:construction}
With the extracted set of triplets, the graph construction is to convert these triplets into a case graph. For a legal case $c$ and its triplet set $R_{c,\text{fact}}$ and $R_{c,\text{issue}}$ specifically for its \textit{legal fact} and \textit{legal issue}, the TACG is constructed as $G_{c,\text{fact}}=(V_{c,\text{fact}}, E_{c,\text{fact}})$ and $G_{c,\text{issue}}=(V_{c,\text{issue}}, E_{c,\text{issue}})$. For $G_{c,\text{fact}}$, $V_{c,\text{fact}}$ includes the set of nodes of all head and tail entities $h$ and $t$ in $R_{c,\text{fact}}$ as $v_h$ and $v_t$, and $E_{c,\text{fact}}$ includes the set of edges corresponding to the relations $r$ from head entity $h$ to tail entity $t$ in $R_{c,\text{fact}}$ as $e_{v_hv_t}$. The same construction process is applied to the $G_{c,\text{issue}}$ for the legal issue. Additionally, for both the legal fact graph and the legal issue graph of a  case, two virtual node $v_{\text{g, fact}}$ and $v_{\text{g, issue}}$ that representing the global textual semantics are added to fact graph and issue graph respectively. To help propagate the global information in the graph representation learning, this virtual global node is connected to every node in the graph. The detailed illustration of the TACG is demonstrated in Fig.~\ref{fig:TACG}.

\subsubsection{Text Attribute.}
\label{sec:ta}
In the case graph above with the extracted entities as nodes and relations as edges, the node and edge features are obtained by using language models encoding of the text in nodes and edges. For node $u$, node $v$, and edge $e_{uv}$ in the case, the text attribute encoding is processed as :
\begin{equation}
\label{eq:TAencoding}
    \mathbf{x}_u = \text{LM}(t_u);\quad
    \mathbf{x}_v = \text{LM}(t_v);\quad
    \mathbf{x}_{e_{uv}} = \text{LM}(t_{e_{uv}}),
\end{equation}
where $t_u$,$t_v$,$t_{e_{uv}}$ is the text of node $u$, node $v$ and edge $e_{uv}$ respectively, and LM is a pre-trained language model, such as BERT~\cite{BERT}, SAILER~\cite{SAILER} or PromptCase~\cite{promptcase}. $\textbf{x}_u \in \mathbb{R}^d$, $\textbf{x}_v \in \mathbb{R}^d$, and $\textbf{x}_{e_{uv}} \in \mathbb{R}^d$ are the output of text-attributed encoding and serve as the feature vector of node $u$, $v$ and edge $e_{uv}$. For the virtual global nodes $v_\text{g,fact}$ and $v_\text{g,issue}$, the node feature is whole text encoding of the legal fact and legal issue extracted by using PromptCase~\cite{promptcase} from a case respectively. The feature of the edge between the virtual gloabal node and other nodes such as entity $u$ will directly reuse the feature of other nodes $u$ in the TACG to simplify the feature extraction:
\begin{equation}
\label{eq:global_text}
\begin{split}
    \mathbf{x}_{v_\text{g,fact}}=\text{LM}(t_\text{fact}); \quad \mathbf{x}_{e_{uv_\text{g,fact}}}=\mathbf{x}_u;\\ 
    \mathbf{x}_{v_\text{g,issue}}=\text{LM}(t_\text{issue}); \quad \mathbf{x}_{e_{uv_\text{g,issue}}}=\mathbf{x}_u.
\end{split}
\end{equation}

\subsection{Edge Graph Attention Layer}
\label{sec:EGAT}
After obtaining the text-attributed features of nodes and edges, a self-attention module will be used to aggregate the nodes and its neighbour nodes and edges information to an informative representation. Moreover, to avoid over-smoothing, a residual connection is added. As previous study shows, multi-head attention has better performance than original attention\cite{transformer}. According to multi-head attention mechanism, the update of node $v$ with $K$ attention heads is defined as:
\begin{equation}
  \mathbf{h}^{'}_v = \mathbf{W}_s\cdot \mathbf{h}_v + \mathop{\text{Avg}}_{k=1:K}(\sum_{u\in N(v)}\alpha^{k}(\mathbf{W}^{k}_n \cdot \mathbf{h}_u+\mathbf{W}^{k}_e \cdot \mathbf{h}_{e_{uv}}),
\end{equation}
where $\mathbf{h}^{'}_v\in\mathbb{R}^{d'}$ is the updated node feature, and Avg means the average of the output vectors of K heads. Specially, the input of the first EdgeGAT layer is initialised as $\textbf{h}_v=\textbf{x}_v$, $\textbf{h}_u=\textbf{x}_u$ and $\textbf{h}_{e_{uv}}=\textbf{x}_{e_{uv}}$. All the weight matrices $\textbf{W}$ are in $\mathbb{R}^{d\times d'}$. Specifically, $\textbf{W}_{s}$ is the node self update weight matrix, $\textbf{W}_{n}$ is the neighbour node update weight matrix and $\textbf{W}_{e}$ is the edge update weight matrix respectively. $\alpha^{k}$ is the attention weight in the attention layer as:
\begin{equation}
    \alpha^{k}= \text{Softmax}(\text{LeakyReLU}({\mathbf{w}_{\text{att}}^{k}}^{T} [\mathbf{W}^{k}_{n} \cdot \mathbf{h}_v\mathbin\Vert \mathbf{W}^{k}_{n} \cdot \mathbf{h}_u \mathbin\Vert \mathbf{W}^{k}_{e} \cdot \mathbf{h}_{e_{uv}}])),
\end{equation}
where $\text{Softmax}$ is the softmax function, $\text{LeakyReLU}$ is the non-linear function, $\mathbf{w}_{\text{att}}^{k} \in \mathbb{R}^{3d'}$ is the weight vector of attention layer and $\mathbin\Vert$ denotes concatenation of vectors. Specifically, the same edge features are reused to make the EdgeGAT simpler. Further model development of updating edge can be designed.  

\subsection{Readout Function}
\label{sec:readout}
With the updated node and edge representations, a graph readout function is designed to obtain the case graph representation for the case $c$:
\begin{equation}
  \mathbf{h}_{c,\text{fact}} = \text{Readout}(G_{c,\text{fact}}),\quad  \mathbf{h}_{c,\text{issue}} = \text{Readout}(G_{c,\text{issue}}),
\end{equation}
where Readout is an aggregation function to output an overall representation in the graph level. One example is the average pooling of all the node embeddings:
\begin{equation}
  \mathbf{h}_{c,\text{fact}} = \text{Avg}(\textbf{h}_{v_{i}}|v_{i} \in V_{c,\text{fact}});\quad  \mathbf{h}_{c,\text{issue}} = \text{Avg}(\textbf{h}_{v_{i}}|v_{i} \in V_{c,\text{issue}}).
\end{equation}

In addition, since the virtual global node has already been in TACG, the updated virtual global node vector $\mathbf{h}_g$ can be considered as the final representation of the case graph:
\begin{equation}
  \mathbf{h}_{c,\text{fact}} = \mathbf{h}_{g,\text{fact}};\quad  \mathbf{h}_{c,\text{issue}} = \mathbf{h}_{g,\text{issue}}.
\end{equation}
For the experiments in this paper, the final virtual global node vector is used as the graph representation.

For case $c$, the fact graph feature $\mathbf{h}_{c,\text{fact}}$ and issue graph feature $\mathbf{h}_{c,\text{issue}}$ are generated by using $\textbf{h}_{c,\text{fact}}$ and $\textbf{h}_{c,\text{issue}}$ respectively. Therefore, the case graph representation $\textbf{h}_c \in \mathbb{R}^{2d'}$ is the concatenation of $\textbf{h}_{c,\text{fact}}$ and $\textbf{h}_{c,\text{issue}}$:
\begin{equation}
  \textbf{h}_c = \textbf{h}_{c,\text{fact}} \mathbin\Vert \textbf{h}_{c,\text{issue}}.
\end{equation}

\subsection{Objective Function}
\label{sec:obj}
To train the CaseGNN model for the LCR task, it is required to distinguish the relevant cases from irrelevant cases given a large candidate case pool. To provide the training signal, contrastive learning is a tool that aims at pulling the positive samples closer while pushing the negative samples far away used in retrieval tasks~\cite{SAILER,MVCL,CL4LJP}. In this paper, given a query case $q$ and a set of candidate case $D$ that includes both relevant cases $d^+$ and irrelevant cases $d^-$, the objective function is defined as a contrastive loss:
\begin{equation}
\label{eq:con}
  \ell = -\text{log}\frac{e^{(s(\mathbf{h}_q,\mathbf{h}_{d^+}))/\tau}}{e^{(s(\mathbf{h}_q,\mathbf{h}_{d^+}))/\tau}+\sum^n_{i=1}e^{(s(\mathbf{h}_q,\mathbf{h}_{d^{easy-}_i}))/\tau}+\sum^m_{j=1}e^{(s(\mathbf{h}_q,\mathbf{h}_{d^{hard-}_j}))/\tau}},
\end{equation}
where $s$ is the similarity metric such as dot product or cosine similarity, $n$ is the number of easy negative samples, $m$ is the number of hard negative samples, and $\tau$ is the temperature coefficient. During training, the positive samples are given by the ground truth from the dataset. The easy negative samples are randomly sampled from the whole candidate pool as well as using the in-batch samples from other queries. For the hard negative samples, it is designed to make use of harder samples to effectively guide the training. Therefore, hard negative samples are sampled based on the BM25 relevance score. If a candidate case has a high score from BM25 yet it is not a positive case, such a case is considered as a hard negative case because it has a high textual similarity to the query case while it is still not a positive case. The overall pipeline is detailed in Fig.~\ref{fig:CaseGNN}.

\section{Experiments}
\subsection{Setup}
\label{sec:setup}
\begin{wraptable}{r}{0.5\linewidth}\centering
    \vspace{-0.7cm}
    \caption{Statistics of datasets.}
    \label{tab:dataset}
    \resizebox{1\linewidth}{!}{
    \begin{tabular}{c|cc|cc}
    \toprule
    \multirow{2}{*}{Datasets} &\multicolumn{2}{c|}{COLIEE2022} &\multicolumn{2}{c}{COLIEE2023} \\
    \cmidrule{2-5}
    &train &test &train &test \\\midrule
    \# Query &898 &300 &959 &319 \\
    \# Candidates &4415 &1563 &4400 &1335 \\
    \# Avg. relevant cases &4.68 &4.21 &4.68 &2.69 \\
    Avg. length (\# token) &6724 &6785 &6532 &5566 \\
    Largest length (\# token) &127934 &85136 &127934 &61965 \\
    \bottomrule
    \end{tabular}
    }
    \vspace{-0.7cm}
\end{wraptable}

\subsubsection{Datasets.}
To evaluate the proposed CaseGNN, the experiments are conducted on two benchmark LCR datasets, COLIEE2022~\cite{COLIEE2022} and COLIEE2023~\cite{COLIEE2023} from the Competition on Legal Information Extraction/Entailment (COLIEE), where the cases are collected from the federal court of Canada. Given a query case, relevant cases are retrieved from the entire candidate pool. The difference between two datasets are: (1) Although the training sets have overlap, the test sets are totally different; (2) As shown in Table~\ref{tab:dataset}, the average relevant cases numbers per query are different, leading to different difficulties in finding relevant cases. These datasets focus on the most widely used English legal case retrieval benchmarks and CaseGNN can be easily extended to different languages with the corresponding information extraction tools and LMs.

\subsubsection{Metrics.}
\label{metrics}
In this experiment, the metric of precision (P), recall (R), Micro F1 (Mi-F1), Macro F1 (Ma-F1), Mean Reciprocal Rank (MRR), Mean Average Precision (MAP) and normalized discounted cumulative gain (NDCG) are used for evaluation. According to the previous LCR works~\cite{promptcase,LeCaRD,SAILER}, top 5 ranking results are evaluated. All metrics are the higher the better.

\subsubsection{Baselines.}
\label{baselines}
According to the recent research~\cite{promptcase,SAILER}, 5 popular and state-of-the-art methods are compared as well as the competition winners:
\begin{itemize}
    \item \textbf{BM25}~\cite{BM25}: a strong retrieval benchmark that leverages both term frequency and inverse document frequency for retrieval tasks.
    \item \textbf{LEGAL-BERT}~\cite{LEGAL-BERT}: a legal LM that is pre-trained on large English corpus.
    \item \textbf{MonoT5}~\cite{monot5}: a pre-trained LM that utilises T5~\cite{T5} architecture for document ranking tasks.
    \item \textbf{SAILER}~\cite{SAILER}: a pre-trained legal structure-aware LM that obtains competitive performance on both datasets. 
    \item \textbf{PromptCase}~\cite{promptcase}: an input reformulation method that works on LM for LCR, which achieves sate-of-the-art performance on COLIEE2023~\cite{COLIEE2023} dataset. Two-stage usage of PromptCase with BM25 is evaluated as well.
\end{itemize}

\subsubsection{Implementation.}
\label{implementation}
The French text in both datasets are removed. The spaCy\footnote{\url{https://spacy.io/}}, Stanford OpenIE~\cite{OpenIE} and LexNLP\footnote{\url{https://github.com/LexPredict/lexpredict-lexnlp}} packages are used for information extraction. Two-stage experiment uses the top 10 retrieved cases by BM25 as the first stage result. The embedding size are set to 768 based on BERT. The number of EdgeGAT layers are set to 2 and the number of EdgeGAT heads are chosen from \{1, 2, 4\}. The training batch sizes are chosen from \{16, 32, 64, 128\}. The Dropout~\cite{dropout} rate of every layer's representation is chosen from \{0.1, 0.2, 0.3, 0.4, 0.5\}. Adam~\cite{Adam} is applied as optimiser with the learning rate chosen from \{0.00001, 0.00005, 0.0001, 0.0005, 0.000005\} and weight decay from \{0.00001, 0.0001, 0.001, 0.01\}. For every query during training, the number of positive sample is set to 1; the number of randomly chosen easy negative sample is set to 1; the number of hard negative samples is chosen from \{1, 5, 10, 30\}. The in-batch samples from other queries are also employed as easy negative samples. SAILER~\cite{SAILER} is chosen as the LM model to generate the text attribute of nodes and edges, which is a BERT-based model that pre-trained and fine-tuned on large corpus of legal cases.

\begin{table}[!t]\centering
\caption{Overall performance on COLIEE2022 and COLIEE2023 (\%). Underlined numbers indicate the best baselines. Bold numbers indicate the best performance of all methods. Both one-stage and two-stage results are reported.}\label{tab:overall}
\resizebox{1\linewidth}{!}{
\begin{tabular}{l|ccccccc|ccccccc}
\toprule
\multirow{2}{*}{Methods} &\multicolumn{7}{c}{COLIEE2022} &\multicolumn{7}{c}{COLIEE2023}\\
\cmidrule{2-15}
&P@5 &R@5 &Mi-F1 &Ma-F1 &MRR@5 &MAP &NDCG@5 &P@5 &R@5 &Mi-F1 &Ma-F1 &MRR@5 &MAP &NDCG@5 \\\midrule
\midrule
\textbf{One-stage}\\
BM25 &\underline{17.9} &\underline{21.2} &\underline{19.4} &\underline{21.4} &23.6 &25.4 &33.6 &\underline{16.5} &\underline{30.6} &\underline{21.4} &\underline{22.2} &23.1 &20.4 &23.7\\
LEGAL-BERT &4.47 &5.30 &4.85 &5.38 &7.42 &7.47 &10.9 &4.64 &8.61 &6.03 &6.03 &11.4 &11.3 &13.6\\
MonoT5 &0.71 &0.65 &0.60 &0.79 &1.39 &1.41 &1.73 &0.38 &0.70 &0.49 &0.47 &1.17 &1.33 &0.61 \\
SAILER &16.6 &15.2 &14.0 &16.8 &17.2 &18.5 &25.1 &12.8 &23.7 &16.6 &17.0 &25.9 &25.3 &29.3\\
PromptCase &17.1 &20.3 &18.5 &20.5 &\underline{35.1} &\underline{33.9} &\underline{38.7} &16.0 &29.7 &20.8 &21.5 &\underline{32.7} &\underline{32.0} &\underline{36.2} \\
\rowcolor{Gray}
CaseGNN (Ours) &\textbf{35.5}$\pm$0.2 &\textbf{42.1}$\pm$0.2 &\textbf{38.4}$\pm$0.3 &\textbf{42.4}$\pm$0.1 &\textbf{66.8}$\pm$0.8 &\textbf{64.4}$\pm$0.9 &\textbf{69.3}$\pm$0.8 &\textbf{17.7}$\pm$0.7 &\textbf{32.8}$\pm$0.7&\textbf{23.0}$\pm$0.5 &\textbf{23.6}$\pm$0.5 &\textbf{38.9}$\pm$1.1 &\textbf{37.7}$\pm$0.8 &\textbf{42.8}$\pm$0.7\\\midrule
\midrule
\textbf{Two-stage}\\
SAILER &\textbf{\underline{23.8}} &\underline{25.7} &\underline{24.7} &25.2 &\underline{43.9} &\underline{42.7} &\underline{48.4} &19.6 &32.6 &24.5 &23.5 &37.3 &36.1 &40.8\\
PromptCase &23.5 &25.3 &24.4 &\textbf{\underline{30.3}} &41.2 &39.6 &45.1 &\textbf{\underline{21.8}} &\underline{36.3} &\textbf{\underline{27.2}} &\underline{26.5} &\underline{39.9} &\underline{38.7} &\underline{44.0}\\
\rowcolor{Gray}
CaseGNN (Ours) &22.9$\pm$0.1 &\textbf{27.2}$\pm$0.1 &\textbf{24.9}$\pm$0.1 &27.0$\pm$0.1 &\textbf{54.9}$\pm$0.4 &\textbf{54.0}$\pm$0.5 &\textbf{57.3}$\pm$0.6 &20.2$\pm$0.2 &\textbf{37.6}$\pm$0.5 &26.3$\pm$0.3 &\textbf{27.3}$\pm$0.2 &\textbf{45.8}$\pm$0.9 &\textbf{44.4}$\pm$0.8 &\textbf{49.6}$\pm$0.8\\
\bottomrule
\end{tabular}}
\end{table}

\subsection{Overall Performance}
\label{overall}
In this experiments, the overall performance of CaseGNN is evaluated on COLIEE2022 and COLIEE2023 by comparing with state-of-the-art models, as shown in Table~\ref{tab:overall}. According to the results, CaseGNN achieves the best performance compared with all the baseline models by a large margin in both one-stage and two-stage settings. In COLIEE2022, one-stage CaseGNN has a much higher performance than other two-stage methods, and in COLIEE2023, one-stage CaseGNN has a comparable results with two-stage methods.

For one-stage retrieval setting, compared to the state-of-the-art performance on COLIEE2022 and COLIEE2023, CaseGNN significantly improved the LCR performance by utilising the important legal structural information with graph neural network. Compared with the strong baseline of traditional retrieval model BM25, CaseGNN achieves outstanding performance. The reason of inferior performance of BM25 is because only using the term frequency will ignore the important legal semantics of a case. The performance of CaseGNN also outperforms LEGAL-BERT, a legal corpus pre-trained LCR model, which indicates that only using legal corpus to simply pre-trained on BERT-based model is not enough for the difficult LCR task. The performances of MonoT5 model on two datasets are the poorest in the experiment, which may for the reason that MonoT5 is pre-trained for text-to-text tasks instead of information retrieval tasks. Although SAILER model uses the structure-aware architecture, the performances are not that good as CaseGNN, which shows that the combining of both TACG and GNN model can largely improve the learning and understanding ability of model. Since CaseGNN model uses the fact and issue format to construct the TACG to encode a graph representation, the worse performances of PromptCase indicates the importance of transforming fact and issue into TACG and applying to GNN to learn an expressive case graph representation for LCR.

For two-stage retrieval setting, all methods use a BM25 top10 results as the first stage retrieval and conduct the re-ranking based on these ten retrieved cases. SAILER and PromptCase are compared since these two methods have a comparable one-stage retrieval performance. CaseGNN outperforms both methods in most metrics, especially those related to ranking results, such as MRR@5, MRR and NDCG@5. In COLIEE2022, although CaseGNN has a much higher performance than other baselines, CaseGNN actually cannot benefit from a two-stage retrieval since BM25 cannot provide a higher and useful first stage ranking result compared with CaseGNN itself. In COLIEE2023, all methods can benefit from the two-stage retrieval and CaseGNN can further improve the performance over the one-stage setting.

\begin{table}[!t]\centering
\caption{Ablation study. (\%)}\label{tab:ablation}
\resizebox{1\linewidth}{!}{
\begin{tabular}{l|ccccccc|ccccccc}
\toprule
\multirow{2}{*}{Methods} &\multicolumn{7}{|c|}{COLIEE2022} &\multicolumn{7}{c}{COLIEE2023}\\
\cmidrule{2-15}
&P@5 &R@5 &Mi-F1 &Ma-F1 &MRR@5 &MAP &NDCG@5 &P@5 &R@5 &Mi-F1 &Ma-F1 &MRR@5 &MAP &NDCG@5 \\
\midrule\midrule
PromptCase &17.1 &20.3 &18.5 &20.5 &35.1 &33.9 &38.7 &16.0 &29.7 &20.8 &21.5 &32.7 &32.0 &36.2 \\
w/o $v_g$ &1.6$\pm$0.1 &2.9$\pm$0.1 &2.1$\pm$0.1 &2.2$\pm$0.1 &4$\pm$0.1 &4.0$\pm$0.1 &4.8$\pm$0.2 &1.6$\pm$0.1 &2.9$\pm$0.1 &2.1$\pm$0.1 &2.2$\pm$0.1 &4.0$\pm$0.1 &2.9$\pm$0.1 &4.8$\pm$0.1 \\
Avg Readout &30.5$\pm$0.5 &36.2$\pm$0.6 &33.1$\pm$0.5 &36.8$\pm$0.5 &61.3$\pm$0.4 &59.0$\pm$0.1 &64.6$\pm$0.5 &17.6$\pm$0.4 &32.6$\pm$0.7 &22.8$\pm$0.5 &23.6$\pm$0.4 &37.7$\pm$1.0 &36.6$\pm$0.7 &41.7$\pm$0.5 \\
\rowcolor{Gray}
CaseGNN &35.5$\pm$0.2 &42.1$\pm$0.2 &38.4$\pm$0.3 &42.4$\pm$0.1 &66.8$\pm$0.8 &64.4$\pm$0.9 &69.3$\pm$0.8 &17.7$\pm$0.7 &32.8$\pm$0.7&23.0$\pm$0.5 &23.6$\pm$0.5 &38.9$\pm$1.1 &37.7$\pm$0.8 &42.8$\pm$0.7\\
\bottomrule
\end{tabular}}
\end{table}

\subsection{Ablation Study}
The ablation study is conducted to verify the effectiveness of the graph components of CaseGNN: (1) not using any graph, which is equivalent to PromptCase; (2) using TACG without the virtual global node (w/o $v_g$); and (3) using the average of updated node features as case graph representation (Avg Readout). The experiments are conducted on both datasets and measured under all metrics. Results are reported in Table~\ref{tab:ablation}. Only one-stage experiments are considered.

As shown in Table~\ref{tab:ablation}, the CaseGNN framework with all the proposed components can significantly outperform the other variants for both datasets. PromptCase utilises the text encodings of the legal fact and the legal issue to obtain the case representation, which serves as a strong baseline for LCR. The virtual global node in TACG comes from the encoding of the legal fact and the legal issue. For the w/o $v_g$ variant, the performance is the worst that the model almost cannot learn any useful information because without the proper text encodings, the overall semantics are not effectively encoded. For the AvG Readout variant, the readout function of CaseGNN is set to using the average node embeddings as the case graph representation. This variant is outperformed by CaseGNN because in this variant, the readout function ignores the information in the edge features. Nevertheless, Avg Readout has a better result compared with PromptCase, which verifies that the graph structure in the case can provide useful information.

\begin{table}[!t]\centering
\caption{Effectiveness of GNNs. (\%)}\label{tab:effect_gnn}
\resizebox{1\linewidth}{!}{
\begin{tabular}{l|ccccccc|ccccccc}
\toprule
\multirow{2}{*}{Methods} &\multicolumn{7}{|c|}{COLIEE2022} &\multicolumn{7}{c}{COLIEE2023}\\
\cmidrule{2-15}
&P@5 &R@5 &Mi-F1 &Ma-F1 &MRR@5 &MAP &NDCG@5 &P@5 &R@5 &Mi-F1 &Ma-F1 &MRR@5 &MAP &NDCG@5 \\
\midrule\midrule
GCN &21.3$\pm$0.3 &25.3$\pm$0.4 &23.2$\pm$0.2 &26.0$\pm$0.4 &46.0$\pm$0.2 &44.4$\pm$0.1 &49.7$\pm$0.3 &12.8$\pm$0.2 &23.7$\pm$0.3 &16.5$\pm$0.1 &16.9$\pm$0.2 &27.8$\pm$0.8 &26.8$\pm$0.6 &31.6$\pm$0.7\\
GAT &29.3$\pm$0.1 &34.8$\pm$0.3 &31.8$\pm$0.1 &35.3$\pm$0.2 &59.2$\pm$0.5 &56.9$\pm$0.3 &62.3$\pm$0.7 &17.4$\pm$0.3 &32.2$\pm$0.5 &22.5$\pm$0.3 &23.1$\pm$0.5 &37.5$\pm$0.4 &36.4$\pm$0.4 &41.4$\pm$0.4  \\
\rowcolor{Gray}
EdgeGAT &35.5$\pm$0.2 &42.1$\pm$0.2 &38.4$\pm$0.3 &42.4$\pm$0.1 &66.8$\pm$0.8 &64.4$\pm$0.9 &69.3$\pm$0.8 &17.7$\pm$0.7 &32.8$\pm$0.7&23.0$\pm$0.5 &23.6$\pm$0.5 &38.9$\pm$1.1 &37.7$\pm$0.8 &42.8$\pm$0.7\\
\bottomrule
\end{tabular}}
\end{table}

\subsection{Effectiveness of GNNs}
To validate the effectiveness of the EdgeGAT layer, CaseGNN is compared with variants of substituting EdgeGAT with GCN~\cite{GCN} and GAT~\cite{GAT}. The experiments are conducted on both datasets and evaluated with all metrics. The results are shown in Table~\ref{tab:effect_gnn}. Only one-stage experiments are considered. For a fair comparison, all variants will be trained with the proposed TACG.

As shown in the experimental results, EdgeGAT has the highest performance compared with the widely used GNN models GCN and GAT. The outstanding results of EdgeGAT on LCR tasks is because in the proposed CaseGNN method, there is a TACG module including both node and edge features in the case graph. Correspondingly, EdgeGAT has a novel design to incorporate the edge features into the case representation calculation. These edge features are important in terms of the legal information contained in the relations between different entities extracted from the legal case. For both GCN and GAT, since these general GNN layers do not have the capability to utilise the edge information, only the node information encoded from the entities are used in the calculation, which leads to information loss of the legal case. More specifically, GAT has a better performance compared with GCN. This phenomenon is aligned with the performance gap between GAT and GCN on other general graph learning tasks because of the graph learning ability difference between GAT and GCN.

\subsection{Parameter Sensitivity}
\begin{figure}[!h]
\centering
    \subfigure{
    \includegraphics[width=0.4\linewidth]{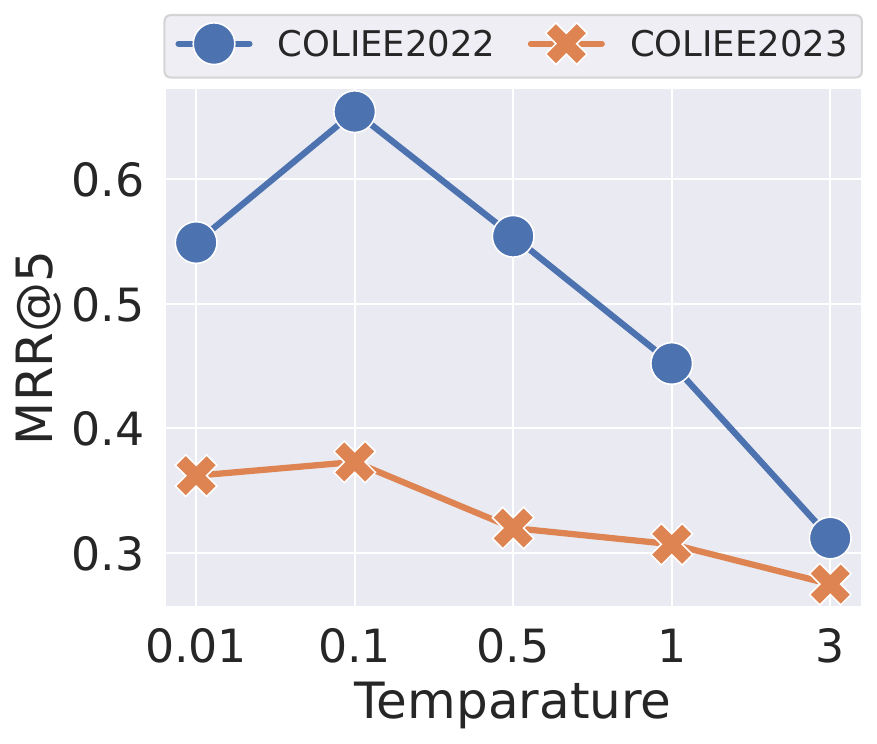}
    \label{fig:temp-mrr}
    }
    \subfigure{
    \includegraphics[width=0.4\linewidth]{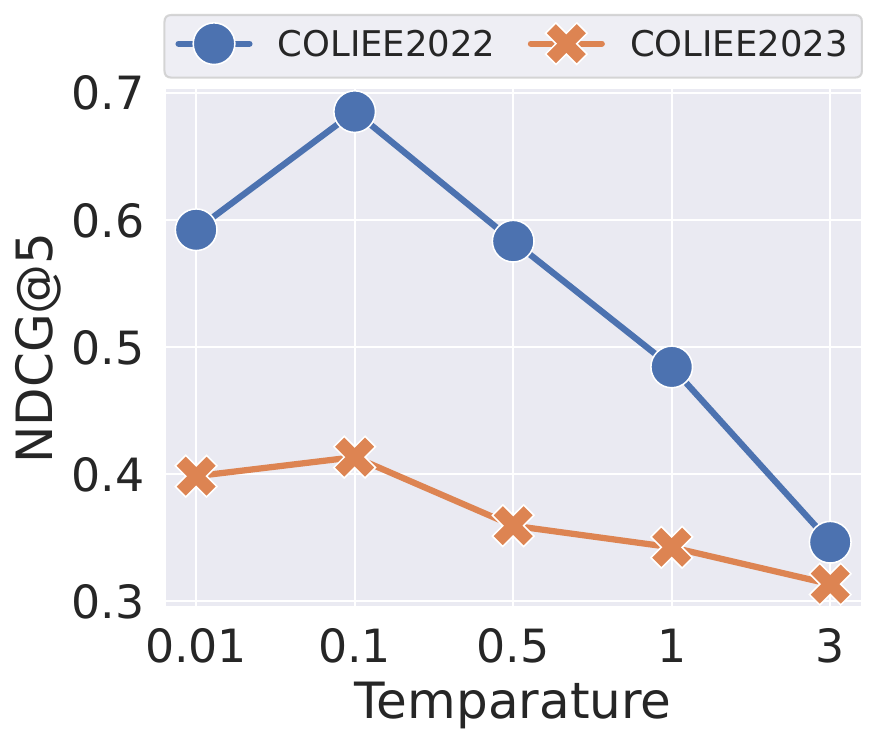}
    \label{fig:temp-ndcg}
    }
\caption{Parameter sensitivity for the temperature $\tau$ in the contrastive loss.}
\label{fig:temp}
\end{figure}

\begin{figure}[!h]
\centering
    \subfigure{
    \includegraphics[width=0.4\linewidth]{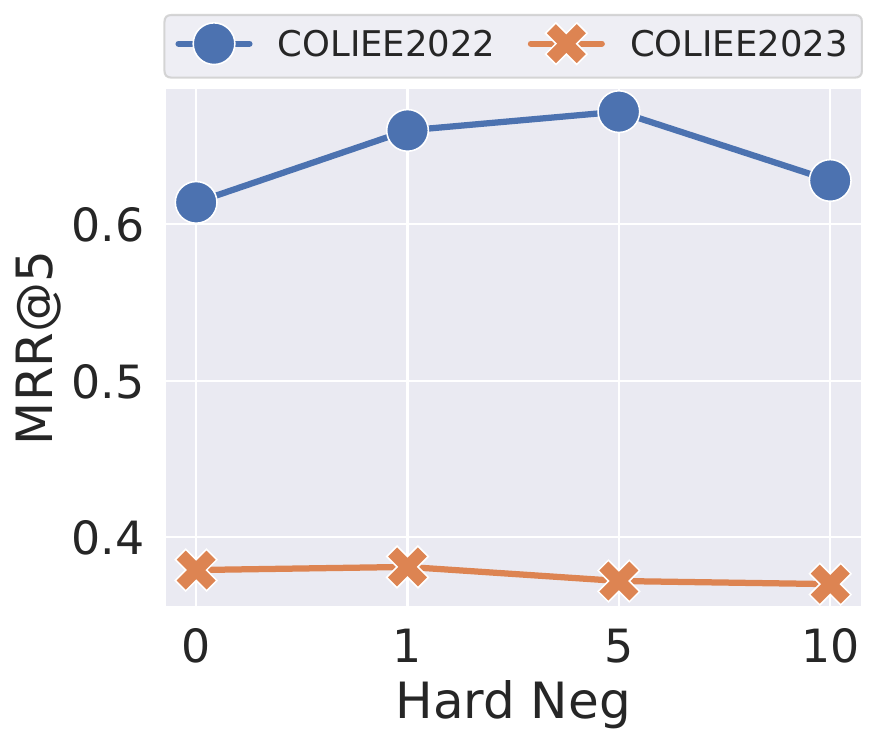}
    \label{fig:neg-mrr}
    }
    \subfigure{
    \includegraphics[width=0.4\linewidth]{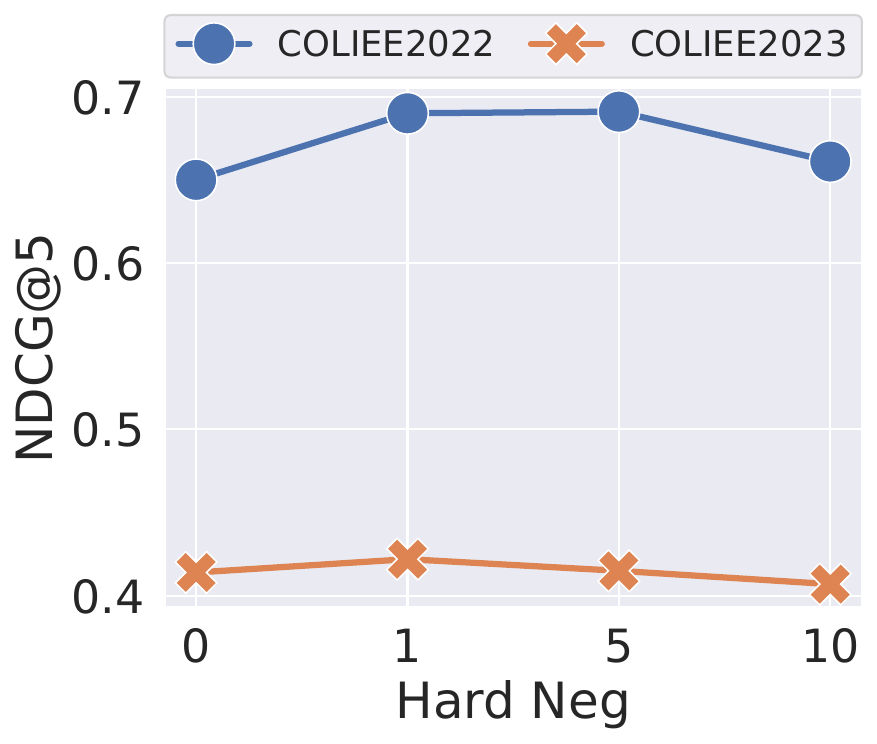}
    \label{fig:neg-ndcg}
    }
    \vspace{-0.5cm}
\caption{Parameter sensitivity for the number of hard negative samples in the contrastive loss.}
\label{fig:hard-neg}
\end{figure}

In this experiment, the temperature coefficient $\tau$ and the number of hard negative samples in the contrastive loss in Equation~(\ref{eq:con}) are investigated for their parameter sensitivity. The results are presented in Fig.~\ref{fig:temp} and Fig.~\ref{fig:hard-neg} respectively.

\paragraph{Temperature coefficient.} As shown in Fig.~\ref{fig:temp}, temperature is chosen from \{0.01, 0.1, 0.5, 1, 3\}. For both datasets, $\tau$ set to 0.1 achieves the best performance. When the temperature is too large, the similarity score will be flatten and it cannot provide sufficient training signal to the model via the contrastive loss. In the contrast, when $\tau$ is too small, it will extremely sharpen the similarity distribution, which will make the objective function neglect the less significant prediction in the output.

\paragraph{Number of hard negative samples.} According to Fig.~\ref{fig:hard-neg}, the choice of number of hard negative samples are from \{0, 1, 5, 10\}. The hard negative samples are sampled from highly rank irrelevant cases by BM25. Different numbers of hard negative samples in Equation~(\ref{eq:con}) have different impacts to the final model.
When there is no hard negative samples in the training objective, the model will be only trained with easy negative samples by random sampling. Model trained without hard negative samples will have an inferior performance compared with a proper selected number of hard negative samples. This is because hard negative samples can provide a more strict supervision signal to train the CaseGNN. The performance decreases when there are too many hard negative samples, which is because the training task becomes extremely difficult and the model can barely obtain useful information from the training signal.

\begin{figure}[!h]
    \centering
    \includegraphics[width=1\linewidth]{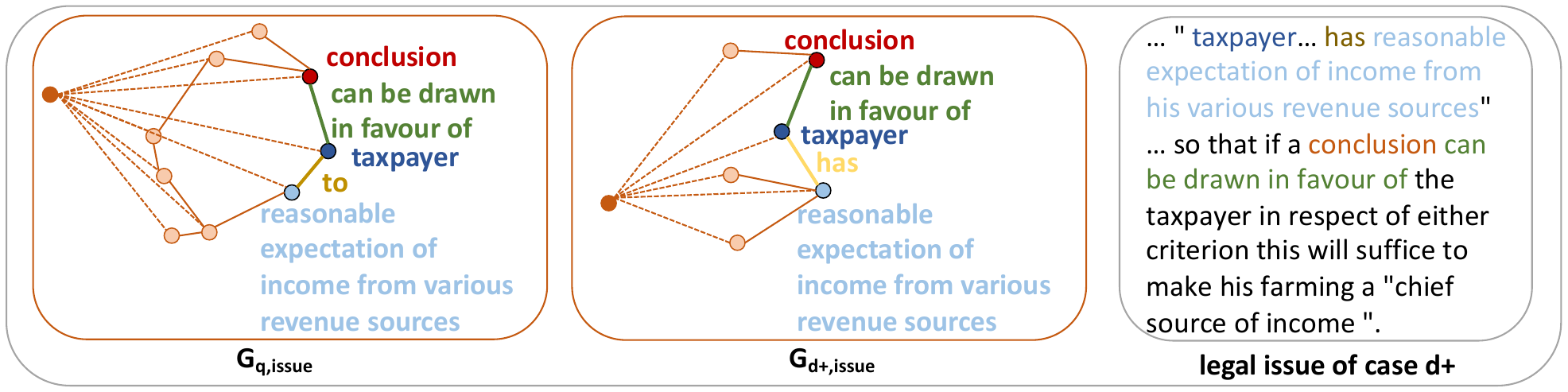}
    \caption{TACGs of cases successfully retrieved by CaseGNN but not by PromptCase.}
    \label{fig:tacg-case}
\end{figure}
\subsection{Case Study}
To further demonstrate how the graph structure helps with the LCR task in CaseGNN, the constructed TACG is visualised in Fig.~\ref{fig:tacg-case}, where CaseGNN successfully performs retrieval while LM-based PromptCase fails. In this visualisation, the constructed TACGs of the legal issue in both the query case $q$ and the ground truth candidate case $d+$, $G_{q,\text{issue}}$ and $G_{d+,\text{issue}}$ are presented. From this visualisation, it it clear that the graph structure can bring multiple entities together to create a candidate graph that is similar to the query graph. On the contrary, from the case text of the candidate $d+$ on the right of Fig.~\ref{fig:tacg-case}, the corresponding language of these entities and relationships are far from each other, which leads to the unsuccessful retrieval of PromptCase for only using sequential LM without case structural information.

\section{Conclusion}
This paper identifies two challenges remaining in recent LM-based LCR models about the legal structural information neglect and the lengthy legal text limitation. To overcome the problems from these challenges, this paper proposes a novel framework, CaseGNN, with specific design to utilise the graph neural networks. A Text-Attributed Case Graph (TACG) is developed to transform the unstructured case text into structural graph data. To learn an effective case representation, an Edge Graph Attention Layer (EdgeGAT) is developed. Extensive experiments conducted on two benchmark datasets verify the state-of-the-art performance and the effectiveness of CaseGNN.

\subsubsection{Acknowledgements} This work is supported by Australian Research Council
CE200100025 and DP230101196.

\bibliographystyle{splncs04}
\bibliography{mybibliography.bib}

\end{document}